\documentclass[preprint,12pt]{aastex}
\usepackage{epsfig, amsmath, amsfonts, amssymb}

\newcommand\bb[1] {   \mbox{\boldmath{$#1$}}  }

\newcommand\del{\bb{\nabla}}
\newcommand\bcdot{\bb{\cdot}}
\newcommand\btimes{\bb{\times}}
\newcommand\vv{\bb{v}}
\newcommand\B{\bb{B}}
\newcommand{\BV}{Brunt-V\"ais\"al\"a\ }

\newcommand\kva{ \bb{k\cdot v_A}  }
\newcommand\bxi{ \bb{\xi} }                
\newcommand\er{  \bb{e_r}  }              
\newcommand\eR{  \bb{e_R}  }

\newcommand\kvA{ \bb{k\cdot v_A}  }

\newcommand{\dd}[2]{\frac{{\rm d} #1}{{\rm d} #2}}

\def\dd{\partial}

\def\beq{ \begin{equation} }
\def\eeq{ \end{equation} }
\def\spose#1{\hbox to 0pt{#1\hss}}  
\def\ltsim{\mathrel{\spose{\lower.5ex\hbox{$\mathchar"218$}}
\raise.4ex\hbox{$\mathchar"13C$}}}
\def\gtsim{\mathrel{\spose{\lower.5ex\hbox{$\mathchar"218$}}
\raise.4ex\hbox{$>$}}}

\slugcomment{Version: \today}

\begin{document}

\title{\bf\LARGE The stability of stratified,
rotating systems and the generation of vorticity in the Sun} 
\author{ Steven A. Balbus\altaffilmark{1,2,3}, 
 Emmanuel Schaan   \altaffilmark{1,3}}

\altaffiltext{1}{Laboratoire de Radioastronomie, \'Ecole Normale
Sup\'erieure, 24 rue Lhomond, 75231 Paris CEDEX 05, France
  \texttt{steven.balbus@lra.ens.fr}}

\altaffiltext{2}{Institut universitaire de France,
Maison des Universit\'es, 103 blvd.\ Saint-Michel, 75005
Paris, France}

\altaffiltext{3}{Department of Astrophysical Sciences, Peyton Hall,
Princeton University, Princeton NJ 08544, USA}

\begin{abstract}

We examine the linear behavior of three-dimensional Lagrangian
displacements in a stratified, shearing background.  The isentropic
and iso-rotation surfaces of the equilibrium flow are assumed to be
axisymmetric, but otherwise fully two-dimensional.  Three-dimensional
magnetic fields are included in the perturbation equations; however the
equilibrium is assumed to be well-described by purely hydrodynamic forces.
The model, in principle very general, is used to study the behavior of
fluid displacements in an environment resembling the solar convection
zone.  Some very suggestive results emerge.  All but high-latitude
displacements align themselves with the observed surfaces of
constant angular velocity.  The tendency for the angular velocity
to remain constant with depth in the bulk of the convective zone,
together with other critical features of the rotation profile, emerge
from little more than a visual inspection of the governing equation.
In the absence of a background axial angular velocity gradient,
displacements exhibit no poleward bias, suggesting that solar convection
``plays-off'' of prexisting shear rather than creates it.  We argue
that baroclinic vorticity of precisely the right order is generated at
the radiative/convective zone boundary due to centrifugal distortion
of equipotential surfaces that is not precisely followed by isothermal surfaces.
If so, many features of the Sun's internal
rotation become more clear, including: i) the general appearance
of the tachocline; ii) the extension of differential rotation well
into the radiative zone; iii) the abrupt change of morphology of convective
zone isorotation surfaces; and iv) the inability of current numerical
simulations to reproduce the solar rotation profile without imposed
entropy boundary conditions.

\end{abstract}

\maketitle



\section{Introduction}

In its most general form, the dynamical state of the interior of
a star is one of differential rotation and entropy stratification.
If isobaric and isochoric surfaces do not coincide, the angular velocity
need not be constant on cylinders.  A noteable example is the Sun, for
which helioseismology studies have fashioned a remarkably detailed and
rich portrait.   Where the Sun is stably stratified in entropy, 
in the bulk of the radiative zone, it tends not to be differentially rotating.
However, surrounding the radius of vanishing entropy gradient, in
both the convective {\em and} radiative layers, there is significant
differential rotation generally dominated by the radial component of
the angular velocity gradient.  Higher in the convection zone, the
rotation contours show an abrupt change in morphology, with the sudden
emergence of a distinctly conical pattern (e.g. Miesch \& Toomre 2009).
Finally, approaching the Sun's surface, there is once again an abrupt
shift in contour morphology, apparently associated with the onset of
high-velocity convection.

In previous work (Balbus, Latter, \& Weiss 2012 [BLW] and references therein),
it has been shown that the pattern of coaxial cones in the bulk of the
solar convective zone (SCZ) can be understood as an elementary solution
of the vorticity equation (in the limit of thermal wind balance) under
certain well-posed assumptions.  One of these assumptions is that a
small angular entropy gradient is present, for without this there can
be no axial component of the angular velocity gradient.  Where does
this all-important entropy gradient come from?  Is it, as is often argued,
an ineluctable consequence of convection and the Coriolis force, or is
something more---or something else---involved?  For that matter, is
the entropy gradient more or less fundamental than the concomitant
angular velocity gradient?  

To address these questions, we begin with a very general study
of the linear behavior of three-dimensional fluid displacements in a
shearing and stratified background medium.
The background angular velocity and entropy profiles may depend upon both
poloidal coordinates.  It is demonstrated that for a medium in uniform
rotation, the most unstable displacements do {\em not} deflect from
spherically radial paths, despite the presence of Coriolis forces.  When a
axial component of the angular velocity gradient is {\em already}
present however, it is shown that there is a significant polar deflection
of higher entropy fluid elements.  More precisely, the sense of this deflection is
poleward for outward-moving displacements if the axial angular velocity
gradient is negative (as in the Sun),
and {\em equatorial} if this gradient is positive.  Thus,
a axial angular angular velocity gradient is self-reinforcing, and may
thus be reshaped, in a convective fluid.  Angular velocity gradients in
cylindrical radius are much less effective in this regard: they have no
first order effect on convective displacements.  Because a background
axial angular velocity gradient requires a vorticity source, this
has far reaching consequences, and we use this finding as a thin edge of
wedge to pry further into the origins of the Sun's baroclinic differential
rotation.  We argue in particular that it is 
likely that the rotation pattern of the SCZ has emerged by responding to a
preexisting angular entropy gradient, rather than generating such a
gradient internally.  This is entirely consistent with the experience of numerical
simulations, in which rotation on cylinders stubbornly persists unless
latitudinal entropy boundary conditions are present, in which case solar-like
profiles emerge relatively easily (e.g. Miesch, Brun,
Toomre 2006).  Indeed, if the conclusions of this paper are well-founded,
the direct imposition of such boundary conditions is 
the ``correct'' procedure!

In the second part of this paper, we put forth the case that vorticity
generation is all but inevitable near the outer edge of the
radiative zone where the entropy gradient vanishes.  The combination of
diffusive heating and centrifugal distortion of equipotential surfaces
is incompatible with radiative and dynamical equilibrium in a uniformly
rotating medium.  The classic remedy of introducing a tiny amount of
meridional circulation (e.g., Schwarzschild 1958) breaks down at a 
surface of zero entropy gradient.  Instead, radiative equilibrium is
re-established with very slightly different isothermal and isochoric
surfaces.  If, as one would expect from radiative considerations, the
isotherms are more spherical than the isochoric surfaces, an incipient
tachocline is generated, bearing many of the features observed of
the true solar tachocline: a negative axial gradient of the angular
velocity everywhere, a dominant (spherical) radial component of this
gradient, and an increasingly dominant cylindrical disposition of the
isorotation contours toward the equator.  

The generation of vorticity at a level stemming from the centrifugal
distortion of the equipotential surfaces has not been hitherto viewed 
as an important component of the solar differential rotation profile.
However, not only is the centrifugal distortion of precisely the correct
order-of-magnitude for this problem, we are argue that it is the principal
causal agent.  If this is correct, numerical simulations whose goal is
to reproduce the Sun's internal rotation accurately from first principles
will ultimately have to accomodate this $1:10^5$ effect.

The organization of the paper is as follows.  In \S 2, we present the
governing equations for three-dimensional fluid displacements in an
axisymmetric but otherwise fully general entropy-stratified,
shearing background.  This is an interesting gasdynamical problem in its own
right, and particularly relevant for the sun.  General solutions
are presented in \S 3, but we focus on the most rapidly growing
modes, for which a simple analysis is possible.   The
solution shows explicitly the relationship between shear, Coriolis
forces, and the deflection of convective trajectories.  In \S 4,
we integrate our findings with standard solar models, arguing that
the seed angular entropy gradient is a result of centrifugal distortion
of equipotential surfaces in the radiative zone together with the 
disappearance of the entropy gradient at the SCZ boundary.  
Finally, \S 5 summarizes our results.

\section{Linear convection theory: fundamental equations}

\subsection{Equilibrium state}

Throughout this paper, we use standard cylindrical coordinates
(with $R$ the radial distance from the rotation axis, $\phi$ the
azimuthal angle, and $z$ the distance along the rotation axis)
and standard
spherical coordinates 
(with $r$ the radial
distance from the origin, $\theta$ the colatitude angle from the $z$
axis, and $\phi$ the azimuthal angle).
Unit vectors will be denoted by an appropriately subscripted $\bb{e}$.

The unperturbed background state is one of time-steady hydrostatic equilibrium,
\beq\label{one}
R\Omega^2\bb{e_R} = {1\over\rho}\del P + \del \Phi 
\eeq
Here, $\Omega$ is the angular velocity, $\rho$ the mass density, $P$ the
gas pressure, and $\Phi$ the gravitational potential.  
The magnetic field $\bb{B}$ is assumed to be weak and
may be ignored in the equilibrium state, but large wavenumber
perturbations could in principle be significantly influenced
by the field.  We therefore will retain the magnetic field when
analyzing small disturbances.
Equation (\ref{one})
is quite general for hydrostatic stars or disks, but for SCZ applications it is an
excellent approximation to take $\Phi
=-GM_\odot/r$ ($G$ is the Newtonian constant and $M_\odot$ is one solar
mass), and to ignore the centrifugal force in the equilibrium state.

The centrifugal term can of course never be ignored in the vorticity equation
(the $\phi$ component of the
curl of equation [\ref{one}]), which provides a measure of the departure
of isobaric and isochoric surfaces:
\beq
R{\dd\Omega^2\over \dd z} =  {1\over\rho^2}\left(
{\dd\rho\over \dd R}{\dd P\over \dd z} -
{\dd\rho\over \dd z}{\dd P\over \dd R} \right),
\eeq
which may also be written
\beq\label{vor}
R{\dd\Omega^2\over \dd z} =  {1\over\gamma \rho}\left(
{\dd\sigma\over \dd z}{\dd P\over \dd R}  -
{\dd\sigma\over \dd R}{\dd P\over \dd z} \right)
\eeq
where $\gamma$ is the adiabatic index, and 
$\sigma\equiv \ln P\rho^{-\gamma}$ is proportional to the
specific entropy.  
The right side of (\ref{vor}) may be written in spherical
coordinates as
\beq\label{vor2}
R{\dd\Omega^2\over \dd z}
=
{1\over r \gamma \rho}\left(
{\dd\sigma\over \dd r}{\dd P\over \dd \theta}  -
{\dd\sigma\over \dd\theta}{\dd P\over \dd r} \right)
\simeq
{g\over r \gamma} {\dd\sigma\over \dd\theta} \quad{\rm (SCZ\ approximation)}
\eeq
where $g=-(1/\rho)(\dd P/\dd r)$ is an excellent approximation to
the gravitational acceleration, and the final approximate equality
assumes that the radial component of the entropy gradient
does {\em not} exceed the latitudinal component by many orders of magnitude.
Equation (\ref{vor2}), often referred to as the thermal wind equation
(e.g. Pedlosky 1987),
appears to be well satisfied throughout much of the
SCZ.  

\subsection {Inertial terms}

To understand more fully the complex dynamics of the SCZ, we analyze here
a much simpler proxy system: the local linear behavior of three-dimensional
linear disturbances in weakly magnetized, stratified, differentially
rotating, two-dimensional background flows.  Note that although the
equilibrium is axisymmetric, the disturbances are fully three-dimensional.
Our focus is the temporal behavior of the fluid displacements embedded
in such a medium.

Nonlinear convection generally involves coherent, extended
structures.  A local WKB treatment cannot hope to capture fully this element
of the problem.  Instead, by affording some insight as to how 
uniformly rotating surfaces arise and host convective displacements,
the local theory can suggest the origin of structure on larger scales.
Such structure is able to survive despite the presence
of shear.

The fundamental dynamical equation of motion for the linear perturbations
is
\beq
{\dd\vv\over\dd t} +(\vv\bcdot\del)\vv +{1\over\rho}\del\left(
P +{B^2\over 8\pi} \right) + \del\Phi - { (\B\bcdot\del)\B\over
4\pi\rho}   = 0.
\eeq
As noted, while the magnetic field $\bb{B}$ is assumed to play no role in the
equilibrium state, it can still be important for the evolution of large
wavenumber perturbations, and will be retained.  We wish to explore the
nonaxisymmetric behavior of local disturbances.  Because the background
flow is in a state of differential rotation, we work in a locally shearing
Lagrangian coordinate system, and the linear perturbations are ultimately
to be expressed in terms of the Lagrangian fluid element displacement
$\bb{\xi} (R, \phi, z, t)$.

Begin by taking standard Eulerian perturbations ($\delta\vv$, $\delta P$, etc.)
of the usual fluid equations. 
The linearized equation of motion is
\beq\label{linear}
{D\ \over Dt} \delta\vv + (\delta\vv\bcdot\del)\vv -{\delta\rho\over\rho^2}
\del P + {\del\over\rho}\left( \delta P + {\B\bcdot\delta \B\over 4\pi}
\right) - { (\B\bcdot\del)\delta \B\over
4\pi\rho}   = 0,
\eeq
where
\beq
{D\ \over Dt} = {\dd \over \dd t} + \Omega{\dd\ \over \dd\phi}
\eeq
is the Lagrangian time derivative associated with the unperturbed flow.
The magnetic pressure buoyancy term and the magnetic tension term
involving the gradient of the background magnetic field have been dropped
under the assumption that the field is weak.
We make the standard WKB assumption that the product of the perturbation
wavenumber with any background scale height is large.

Consider the first two terms of equation (\ref{linear}),
\beq
{D\ \over Dt} \delta\vv + (\delta\vv\bcdot\del)\vv\equiv {\dd\delta\vv\over \dd t}
+\Omega{\dd\delta\vv\over\dd\phi} +(\delta\vv\bcdot\del)(R\Omega\bb{e_\phi})
\eeq
The $R$ and $\phi$ components of these terms are, respectively,
\beq\label{R1}
\bb{e_R}\bcdot \left[
{D\ \over Dt} \delta\vv + (\delta\vv\bcdot\del)\vv\right] 
=
{D\delta v_R\over Dt} -2\Omega\, \delta v_\phi, 
\eeq
\beq\label{phi1}
\bb{e_\phi}\bcdot \left[
{D\ \over Dt} \delta\vv + (\delta\vv\bcdot\del)\vv\right]
=
{D\delta v_\phi \over Dt} +{\kappa^2\over 2\Omega} \delta v_R +
R\left(\dd\Omega\over \dd z\right)\delta v_z, 
\eeq
where 
\beq
\kappa^2 = 2\Omega\left[\Omega + {\dd(R\Omega)\over\dd R}\right].
\eeq
The $z$ component is simply 
\beq
\bb{e_z}\bcdot \left[
{D\ \over Dt} \delta\vv + (\delta\vv\bcdot\del)\vv\right]
=
{D\delta v_z\over Dt}.  
\eeq

The relationship between $\bb{\xi}$ and $\delta \vv$ is given by
\beq\label{dxidt}
{D\bb{\xi}\over Dt} = \delta\vv + \bb{\xi}\bcdot\del(R\Omega\bb{e_\phi})
\eeq
For $j$ equal to $R$ or $z$, we find $\delta v_j=D\xi_j/Dt$, while the
$\phi$ component of (\ref{dxidt}) gives
\beq
{D\xi_\phi\over Dt} = \delta v_\phi + R(\bb{\xi}\bcdot\del)\Omega
\eeq
Equations (\ref{R1}) and (\ref{phi1}) may now be expressed in terms of
$\xi_R$ and $\xi_\phi$, becoming respectively
\beq
\bb{e_R}\bcdot \left[
{D\ \over Dt} \delta\vv + (\delta\vv\bcdot\del)\vv\right]
=\ddot\xi_R - 2\Omega \dot\xi_\phi + R(\bb{\xi}\bcdot\del)\Omega^2,
\eeq
\beq
\bb{e_\phi}\bcdot \left[
{D\ \over Dt} \delta\vv + (\delta\vv\bcdot\del)\vv\right]
=
\ddot\xi_\phi +2\Omega\dot\xi_R,
\eeq
where the dot notation indicates the Lagrangian derivative $D/Dt$.
Putting the last two equations together with the 
$z$ equation of motion leads to the vector equation
\beq
{D\ \over Dt} \delta\vv + (\delta\vv\bcdot\del)\vv = 
\bb{\ddot\xi} + 2\bb{\Omega\times \dot\xi} +\bb{e_R} R(\bb{\xi\cdot}\del)\Omega^2
\eeq
The second term on the right is obviously the Coriolis term,
and the final term is the ``residual centrifugal force:''
the difference between the centrifugal force in the rotating
frame and the forces maintaing the differential rotation.  
Notice the appearance of $\del\Omega^2$, as opposed 
to an angular momentum gradient, as part of the inertial forces.  

\subsection{Linear perturbations: magnetic induction and entropy constraints}

Next, recall the relationship between $\delta \B$
and the displacement $\bb{\xi}$, which follows from the integrated
form of the induction equation,
\beq\label{curl}
\delta\B = \del\btimes (\bb{\xi}\btimes\B)
\eeq
We work in the Boussinesq limit,
\beq
\del\bcdot\bb{\xi} = 0,
\eeq
so that equation (\ref{curl}) becomes 
\beq
\delta\B =(\B\bcdot\del)\bb{\xi}.
\eeq
We have dropped the term $(\bb{\xi}\bcdot\del)\B$ under the WKB assumption that
the displacements are rapidly varying in space.

Finally, for adiabatic perturbations,
\beq
\gamma {\delta\rho\over \rho}  = \bb{\xi}\bcdot\del\sigma,
\eeq
since in the Boussinesq limit, the relative pressure perturbation
$\delta P/P$ is small compared
with the relative density perturbation $\delta\rho/\rho$.  

\subsection {Comoving coordinates and wavenumbers}

\subsubsection {Time dependence of Eulerian wavenumbers}

The final step is to transform to spatial Lagrangian
coordinates comoving with the unperturbed flow, the
``primed'' coordinate system.
This is accomplished by the transformation 
\beq
R'=R, \quad \phi' = \phi -\Omega t, \quad z'=z, \quad t'=t,
\eeq
where of course $\Omega$ is a function of $R$ and $z$.  
Use of the Lagrangian derivative $D/Dt$ ($\equiv \dd/\dd t'$)
has already effected
the transformation of the partial time derivative,
and the two altered poloidal spatial derivatives are
\beq
{\dd\ \over \dd R}= {\dd\ \over \dd R'} -t {\dd\Omega\over \dd R}
{\dd\ \over\dd\phi'}, 
\eeq
\beq
{\dd\ \over \dd z }= {\dd\ \over \dd z'} -t {\dd\Omega\over \dd z}
{\dd\ \over\dd\phi'}.  
\eeq
More compactly,
\beq
\del = \del' -(t\del\Omega)
{\dd\ \over\dd\phi'},
\eeq
a relation that holds for all three components of the gradient.  
In the WKB limit, all disturbances in Lagrangian comoving coordinates
have the spatial dependence
\beq
\exp\left[i\left(k'_R R'  +m\phi'+k'_z z'\right)\right]
\eeq
in which all components of the $\bb{k'}$ wave vector are constants.
This means that the Eulerian spatial derivatives of $R$
and $z$ are replaced locally (and respectively) by $ik_R(t)$ and $ik_z(t)$, where
\beq
k_R(t) = k'_R -mt{\dd\Omega\over \dd R},
\eeq
\beq
k_z(t) = k'_z -mt{\dd\Omega\over \dd z}
\eeq
Henceforth, the time dependence of $k_R$ and $k_z$ will be understood
with
\beq
\dot k_R = -m{\dd\Omega\over \dd R}, \qquad
\dot k_z = -m{\dd\Omega\over \dd z}.   
\eeq
Note that for initially purely azimuthal $e^{im\phi'}$ disturbances,
the poloidal wavenumber components comprise a two-dimensional vector
parallel to $-\del\Omega$.

\subsubsection{Lagrangian displacements and isorotation surfaces}

At sufficiently large times $t$ (or all times if $k'_R=k'_z=0$),
\beq\label{kRkz}
{k_R\over k_z} \rightarrow {\dd\Omega/\dd R\over \dd\Omega/\dd z}
\eeq
and the wave vector becomes increasingly axisymmetric as the poloidal
components grow.  If the three components of the displacement $\bb{\xi}$
are of comparable magnitude, then the condition $\bb{k\cdot\xi}=0$ becomes,
at sufficiently large $t$,        
\beq
 \bb{\xi}\bcdot\del\Omega = 0,  
\eeq
whence
\beq
\bb{\dot\xi}\bcdot\del\Omega = 0.  
\eeq
In other words, {\em independently of the details of the dynamics,} the 
velocity vector of a disturbed fluid element must eventually lie
in a surface of constant
$\Omega$.  
There is nothing ``solar'' about this argument, it relies
entirely on the kinematics of differential rotation and mass conservation
in an incompressible fluid.  But it is tempting
to apply this to the Sun, 
since the fluid elements in question would then be convecting heat
and eliminating excess entropy gradients within surfaces
of constant $\Omega$, and the confluence of these surfaces
with constant residual entropy surfaces becomes less mysterious.
There is a further benefit to such an approach: 
the preponderance of the
constant $\Omega$ surfaces are {\em observed} to be
quasi-radial spokes, as seen in merdional cross section.
They are customarily if crudely described as
cones of constant $\theta$.  If the radially convecting elements
are compelled to follow paths of constant $\Omega$,
then $\Omega\simeq\Omega(\theta)$ is hardly mysterious.
(Precisely the same vanishing-divergence 
reasoning also leads to the conclusion that 
the poloidal components of the magnetic field vector should
lie in constant $\Omega$ surfaces.)

Is this kinematical argument for the alignment of constant residual
entropy and angular velocity surfaces correct?   This depends upon whether
there is sufficient time for the shear to shape the wavelet before 
coherence is lost.  A potential difficulty is that, as the differential
rotation is not large, this may be a rather long time interval depending
upon the initial poloidal wavenumber components.  Long-lived, coherent,
nonlinear and nonlocal structures are seen in fully-developed convection,
and these will in fact tend to lie in constant $\Omega$ surfaces.
In some sense this backs our approach, but at the same time it puts
the cart before the horse:  it is just this rotation-entropy link we
would like to understand.  We suggest that a more rapid linear dynamical
explanation is also available, whereby modes with very small initial
poloidal wavenumbers and near radial displacements are preferred for
rapid development.  This is discussed in more detail in \S 2.6 and \S 3.

\subsubsection{The constancy of $\bb{k\cdot B}$}

The $\phi$ component of the equilibrium magnetic field is not independent
of time, but satisfies the induction equation
\beq
{\dd B_\phi\over \dd t} = R(\B\bcdot\del)\Omega,
\eeq
or 
\beq
B_\phi = B'_\phi + tR(\B\bcdot\del)\Omega.
\eeq
Note, however, that the magnetic tension
$\bb{k\cdot B}=\bb{k'\cdot B'}$, where $\bb{B'}$ is
the magnetic field at $t=0$, is {\em independent of time.}
Introducing the Alfv\'en velocity 
\beq
\bb{v_A} \equiv {\B\over \sqrt{4\pi \rho}},
\eeq
the quantity $(\kva)^2$, the local magnetic tension
force per unit mass, may be regarded as 
a locally constant parameter in the equations of motion.

\subsection{Final Dynamical Equations}

We may now assemble the three fundamental dynamical equations
of motion:
\beq\label{xiR}
\ddot\xi_R -2\Omega \dot\xi_\phi +R(\bb{\xi}\bcdot\del)\Omega^2
- {\dd P\over \dd R}{(\bb{\xi}\bcdot\del) \sigma
\over\gamma\rho}+i
k_R \left( {\delta P\over \rho} + {\bb{B\cdot\delta B}\over 4\pi\rho}
\right) + (\bb{k\cdot v_A})^2\xi_R=0,
\eeq
\beq\label{xiphi}
\ddot\xi_\phi +2\Omega \dot\xi_R 
+{im\over R}
\left( {\delta P\over \rho} + {\bb{B\cdot\delta B}\over 4\pi\rho}
\right) + (\bb{k\cdot v_A})^2\xi_\phi=0,
\eeq
\beq\label{xiz}
\ddot\xi_z 
- {\dd P\over \dd z}{(\bb{\xi}\bcdot\del) \sigma
\over\gamma\rho}+i
k_z \left( {\delta P\over \rho} + {\bb{B\cdot\delta B}\over 4\pi\rho}
\right) + (\bb{k\cdot v_A})^2\xi_z=0.
\eeq
The three dynamical equation may be combined into
a single vectorial equation
\beq\label{legrand}
\ddot{\bxi} + 2\bb{\Omega\times}\dot{\bxi} + \eR R(\bxi\bcdot\del)\Omega^2
-{\del P\over\gamma \rho} \bxi\bcdot\del\sigma  + (\kva)^2\bxi
+ {i\bb{k}\over\rho} \left(\delta P +{\bb{B\cdot\delta B}\over 4\pi}\right)=0
\eeq
with the understanding that $\bb{k}$ in the final term is time dependent,
\beq
\bb{k}(t) = \bb{k}(0) - mt\del\Omega,
\eeq
and 
\beq
\bb{k(t)\cdot\xi} = 0.
\eeq

\subsection {Self-consistent radial convection}

It is a curious and significant fact that in the bulk of the convective
zone the Sun tends to eliminate every extraneous {\em nonconvective} term
in the linear equation (\ref{legrand}), either by the term vanishing
identically or by its cancellation with another nonconvective term.
The dominant dynamics is due almost entirely to the underlying radial
forcing from the unstable entropy gradient, even in the presence of
rotation.

Recall the concept of residual entropy introduced by Balbus et al.\
(2009): the entropy $\sigma(r,\theta)$ is written as the sum of a
function depending only upon spherical radius, $\sigma_r(r)$, and a
residual term, $\sigma' (r,\theta)$.  Physically, $\sigma_r$ represents
the underlying convection-driving unstable radial entropy profile,
and $\sigma'$ is the $\theta$-dependent modification that results as
a consequence of rotation plus convection.  In numerical simulations,
this breakdown has an operational significance: $\sigma_r$ is externally
imposed, and $\sigma'$ is, in essence, the computed reponse (Miesch et
al. 2006) \footnote{In practice, the computed $\sigma'$ may acquire a
purely spherical contribution as well, but this is easily removed by
subtracting off the mean.}.  Only $\sigma'$ is relevant to the thermal
wind equation (\ref{vor2}), since the entropy appears exclusively in
the form of $\dd\sigma/\dd\theta$.

We write
$\sigma=\sigma_r + \sigma'$ and let us assume that $P$ is a function of $r$
only.  Then with $g\er = -(1/\rho) \del P$, the usual squared \BV frequency
$N^2$ is
\beq
N^2 \equiv {g\over\gamma } {\dd\sigma\over \dd r} =
 {g\over\gamma } {\dd(\sigma_r+\sigma')\over \dd r} 
\equiv N_r^2 +{g\over\gamma}
{\dd\sigma'\over \dd r}.
\eeq
In the SCZ, $N^2<0$.
Dropping the magnetic terms, as they appear to be genuinely
tiny, equation (\ref{legrand}) may then be written
\beq\label{legrandr}
\ddot{\bxi} +N_r^2\xi_r \er
+ 2\bb{\Omega\times}\dot{\bxi} + \eR R(\bxi\bcdot\del)\Omega^2
-{\del P\over\gamma \rho} \bxi\bcdot\del\sigma'  
+ {i\bb{k}\delta P\over\rho}  =0
\eeq

Recall that the bulk of the SCZ is characterized by an
angular velocity that is insensitive to depth; roughly speaking,
$\Omega\simeq\Omega(\theta)$.   Moreover, surfaces of constant $\Omega$
and $\sigma'$ coincide well (Miesch et al. 2006; Balbus et al. 2009).  Under these
circumstances, for the dominant radial $\er$ convective
displacements, {\em all terms in equation (\ref{legrandr}) vanish,
cancel hydrostatically, or are otherwise negligible,} save the first
and second.  In particular, 
the Coriolis deflection (third term in from the left)
is balanced by the azimuthal pressure gradient (final term on the left
side),
and the two $\bb{\xi\cdot\nabla}$ terms are are intrinsically small.  
The radially moving disturbances disturbances are characterized by 
poloidal wavenumber components very small compared with $m/R$.
In other words, within the context of simple linear theory
in a uniformly rotating sphere, we have
a plausible beginning for understanding why the Sun's gross pattern
of {\em differential rotation} is the way it is: the most efficient way to
convect heat outwards is by maintaining radial convection, which is
however 
permitted only to the extent that the dynamical forces of differential
rotation allow it.  With $\Omega=\Omega(\theta)$, and fluid motions
embedded
within coinciding $\Omega$ and $\sigma'$ surfaces, the dynamics of
radial covection in a shearing system is self-consistent.  In fact,
the data show that in the bulk of the SCZ, constant $\Omega$ surfaces
are slightly more axial than constant $\theta$ surfaces.
In the next section, we will see that poleward trajectory deviations 
emerge from the solutions of equation (\ref{legrandr}) when
$\dd\Omega/\dd z <0$.  We shall argue, moreover, that these axial
departures from radial trajectories furnish an important clue to
the origin of the Sun's vorticity.

\subsection {Reduction to Two Coupled Equations}

The equation of mass conservation $\del\bcdot\bb{\xi}=0$ may be written
\beq\label{xiphi2}
\xi_\phi =-{R\over m} \left(k_R\xi_R +k_z\xi_z\right).
\eeq
From this it follows
\beq\label{xiphidot}
\dot\xi_\phi = -{R\over m} \left( k_R\dot\xi_R + k_z\dot\xi_z\right)
+R(\bb{\xi}\bcdot\del)\Omega, \qquad
\ddot\xi_\phi = -{R\over m} \left( k_R\ddot\xi_R + k_z\ddot\xi_z\right)
+2R(\bb{\dot\xi}\bcdot\del)\Omega.
\eeq

Next, from equation (\ref{xiz}),
\beq\label{deltaP}
i \left( {\delta P\over \rho} + {\bb{B\cdot\delta B}\over 4\pi\rho}
\right) = -{1\over k_z}
\left[\ddot\xi_z + (\kva)^2\xi_z -
{\dd P\over \dd z}{(\bb{\xi}\bcdot\del) \sigma
\over\gamma\rho}\right]
\eeq
Substituting equations (\ref{xiphi2})- (\ref{deltaP}) into (\ref{xiR})
and (\ref{xiphi}) and simplifying produces the equations
\beq\label{I}
\ddot\xi_R +(\kva)^2\xi_R -{k_R\over k_z}\left(\ddot\xi_z +(\kva)^2\xi_z
\right)
+{2\Omega R\over m}\left( k_R\dot\xi_R+k_z\dot\xi_z\right)
+{ {\cal D}P\over\rho\gamma} (\bb{\xi}\bcdot\del) \sigma
=0,
\eeq
$$
{Rk_R\over m} \left[\ddot\xi_R +(\kva)^2\xi_R\right]
+\left( {Rk_z\over m}+{m\over Rk_z}\right)
\left[\ddot\xi_z +(\kva)^2\xi_z\right]
\qquad\ \ \ 
\qquad\ \ \ 
\qquad\ \ \ 
\qquad\ \ \ 
\qquad\ \ \ 
\qquad\ \ \ 
$$
\beq\label{II}
\qquad
\qquad
\qquad
\qquad
\qquad
\qquad
\qquad
-2\bb{\dot\xi\cdot\nabla}(R\Omega) -
{m\over Rk_z\rho\gamma}
{\dd P\over \dd z}
{ (\bb{\xi}\bcdot\del)} \sigma
=0,
\eeq
where (Balbus 1995):
\beq\label{Deq}
{\cal D} = \left( {k_R\over k_z}{\dd\ \over\dd z}-{\dd\ \over\dd R}\right)
\eeq
Finally, we may recombine equations (\ref{I}) and (\ref{II}), 
separately isolating
$\ddot\xi_R$ and $\ddot\xi_z$:
$$
\ddot\xi_R +(\kva)^2\xi_R
-{2mk_R\over Rk^2}\bb{\dot\xi\cdot\nabla}(R\Omega)
+{2\Omega R\over m}{k_\perp^2\over k^2}\left(k_R\dot\xi_R+k_z\dot\xi_z\right)
\qquad \ \ 
\qquad \ \ 
\qquad \ \ 
\qquad \ \ 
$$
\beq\label{ddotxir}
\qquad \ \ 
\qquad \ \ 
\qquad \ \ 
\qquad \ \ 
\qquad \ \ 
\qquad \ \ 
\qquad \ \ 
\qquad \ \ 
+{1\over\gamma\rho}(\bb{\xi\cdot\nabla}\sigma)\left( {k_z^2\over k^2}{\cal D}P
-{m^2\over R^2k^2}{\dd P\over \dd R} \right) =0,
\eeq
where $k_\perp^2=k_z^2 +m^2/R^2$, and 
$$
\ddot\xi_z + (\kva)^2\xi_z
-{2mk_z\over Rk^2}\bb{\dot\xi\cdot\nabla}(R\Omega)
-{2\Omega R\over m}{k_Rk_z\over k^2}\left(k_R\dot\xi_R+k_z\dot\xi_z\right)
\qquad \ \
\qquad \ \
\qquad \ \
\qquad \ \
$$
\beq\label{ddotxiz}
\qquad \ \
\qquad \ \
\qquad \ \
\qquad \ \
\qquad \ \
\qquad \ \
\qquad \ \
-{1\over\gamma\rho}(\bb{\xi\cdot\nabla}\sigma)\left( {k_Rk_z\over k^2}{\cal D}P
+{m^2\over R^2k^2}{\dd P\over \dd z} \right) =0.   
\eeq
Equations (\ref{ddotxir}) and (\ref{ddotxiz}) are the fundamental coupled
equations governing the behavior of Lagrangian displacements.

\subsection {Plane wave limits: axisymmetry and uniform rotation}
\subsubsection{Axisymmetry}

It is important to establish the axisymmetric behavior of the
disturbances, since it represents the long time behavior of the
nonaxisymmetric reponse.  In particular, we have already noted that at
large times the wavenumber ratio $k_R/k_z$ is just the time-steady ratio
of the corresponding $\Omega$ gradients.  By contrast, $m$ remains fixed,
so the mode becomes asymptotically axisymmetric as time increases and
the poloidal wavenumbers grow.  Thus, at late times {\em all} nonaxisymmetric
modes behave as an axisymmetric mode whose value for $\kvA$ is fixed but
arbitary, while the value for $k_R/k_z$ is fixed by equation (\ref{kRkz}).

In the Appendix, the $m\rightarrow 0$ limit of equations (\ref{ddotxir})
and (\ref{ddotxiz}) is shown to lead to the dispersion relation
of Balbus (1995):      
\beq\label{fifty2}
{k^2\over k_z^2}\varpi^4 +\varpi^2\left[ {1\over R^3}{\cal D}(R^4\Omega^2)
+{1\over\rho\gamma}({\cal D}P)(  {\cal D} \sigma)\right] -
4\Omega^2(\kva)^2 = 0, \qquad \varpi^2 =\omega^2 -(\kva)^2.
\eeq
or
$$
{k^2\over k_z^2}\omega^4+\omega^2\left[
{1\over R^3}{\cal D}(R^4\Omega^2)
+{1\over\rho\gamma}({\cal D}P)(  {\cal D} \sigma)
-2{k^2\over k_z^2}(\kva)^2 \right]
\ \ \ \ \ \ \ \ \ \ \ \ \ \ \ \ \ \ \ \ \ \ \ \ \ \
$$
\beq\label{bigdisp}
\ \ \ \ \ \ \ \ \ \ \ \ \ \ \ \ \ \ \ \ \ \ \ \ \ \
+{k^2\over k_z^2}(\kva)^4 -(\kva)^2
\left[R{\cal D}\Omega^2
+{1\over\rho\gamma}({\cal D}P)(  {\cal D} \sigma)
\right]=0
\eeq

For applications to the SCZ, we are interested in the case in which
the dominant balance of (\ref{bigdisp}) is between the first two terms,
and the magnetic terms are unimportant (Goldreich \& Schubert 1967):
\beq\label{GS}
{k^2\over k_z^2}\omega^2+
{1\over R^3}{\cal D}(R^4\Omega^2)
+{1\over\rho\gamma}({\cal D}P)(  {\cal D} \sigma)
=0
\eeq

With $k_R/k_z=(\dd\Omega/\dd R)(\dd\Omega/\dd z)^{-1}$,
we find that
\beq\label{R4}
{\cal D}(R^4\Omega^2)=-4\Omega^2,
\eeq
and\footnote{The $R$ partial derivative is of course always taken with $z$ constant,
and vice-versa; the $r$ and $\theta$ partial derivatives bear a 
similar relationship.}
\beq
{\cal D}P = \left(\dd\Omega\over \dd z\right)^{-1}
\left( {\dd\Omega\over \dd R}{\dd P\over \dd z}- 
 {\dd\Omega\over \dd z}{\dd P\over \dd R}
\right)
=\left(\dd\Omega\over \dd z\right)^{-1}
\left( \del P\btimes\del\Omega\right)\bcdot\bb{e_\phi}
=
{\dd P\over \dd r}
\left(\dd\Omega\over \dd z\right)^{-1}{1\over r}
{\dd\Omega\over \dd\theta}.
\eeq
Hence, in terms of the gravitational field $g=-(1/\rho)({\dd P/\dd r})$,
\beq\label{mark1}
{1\over\rho\gamma}{\cal D}P = -{g \over \gamma r}
\left(\dd\Omega\over \dd z\right)^{-1}
{\dd\Omega\over \dd\theta}.
\eeq
The right side of this equation consists of directly observed
or easily calculated quantities.  By similar reasoning,
\beq\label{mark2}
{\cal D}\sigma = {\cal D}[\sigma_r(r) + \sigma']=
{d\sigma_r(r) \over d r}
\left(\dd\Omega\over \dd z\right)^{-1}{1\over r}
{\dd\Omega\over \dd\theta}
\eeq
Here we have used the fact ${\cal D}\sigma'=0$, since $\sigma'$ shares
isosurfaces with $\Omega$.   Combining equations (\ref {GS}),
(\ref{R4}), (\ref{mark1}) and (\ref{mark2}), we obtain
\beq
\omega^2 = \left|\del\Omega\right|^{-2}
\left[ \left(\dd\Omega\over\dd z\right)^2 4\Omega^2 +
\left(\dd\Omega\over\dd\theta\right)^2 {g\over \gamma r^2}{d\sigma_r
\over dr}\right]
\eeq
This dispersion relation is an interesting blend,
melding a standard form for inertial/gravity
waves in a {\em uniformly} rotating medium with
$\bb{k}\propto\del\Omega$ for the poloidal wavenumber components,
which obviously requires the presence of
differential rotation to be sensible.  The dynamical effects of the
rotational shear are lost (${\cal D}\Omega=0$) 
when $\bb{k}$ is parallel to $\del\Omega$.  

For the interesting case of $\Omega=\Omega(\theta)$, the dispersion
relation is 
\beq\label{asymp}
\omega^2 = 4\Omega^2\sin^2\theta + {g\over \gamma}{d\sigma_r
\over dr}
\eeq
In principle, 
these axisymmetric modes can be rotationally stabilized
at equatorial or possibly significantly
higher latitudes, depending on how the adverse radial entropy 
gradient is modeled.   If present, this stabilization is of some practical
importance in models in which the Sun is convecting in surfaces of constant
$\Omega$: these asymmetric modes are {\em not} unstable.   Non-axisymmetry
is more than a complication, it is crucial to the convection process itself. 

\subsubsection{Uniform rotation limit}
The dispersion relation for nonaxisymmetric plane waves in the limiting case of a 
uniformly rotating medium may be derived from equations (\ref{ddotxir}) and 
(\ref{ddotxiz}):
\beq
\varpi^4 +\varpi^2\left[ \left(
{1\over\rho\gamma}{\cal D}P\   {\cal D}\sigma - 4\Omega^2 \right)
{k_z^2\over k^2}+ {m^2\over k^2 R^2\gamma\rho}\del P\bcdot\del\sigma
\right] -
4{k_z^2\over k^2}\Omega^2(\kva)^2 = 0
\eeq
where $\varpi^2$ is defined in equation (\ref{fifty2}).  Restricting the
discussion to the nonmagnetic
subscase, the dispersion relation becomes (Cowling 1951):
\beq
\omega^2 + \left[ \left(
{1\over\rho\gamma}{\cal D}P\   {\cal D}\sigma - 4\Omega^2 \right)
{k_z^2\over k^2}+ {m^2\over k^2 R^2\gamma\rho}\del P\bcdot\del\sigma
\right] =0.
\eeq
Notice that $m$ introduces its own form of coupling to the entropy
gradients.  Moreover, it is possible to eliminate rotation-induced moderating
influences on the growth rate by first setting $k_z=0$, and then to 
maximize this rate by setting $k_R=0$.  The maximum growth rate is the
magnitude of the \BV frequency $|N|$, corresponding to precisely radial
$r$ displacements.  The pure $m$ modes are more unstable than are modes
contaminated by poloidal wavenumber components.

How is it possible for an element
to move along a spherical radius in a rotating system
without encountering Coriolis deflections?  The answer, as noted
earlier, is
by striking a geostrophic
balance\footnote{Perhaps ``heliostrophic'' is a more apt description for
the problem at hand.} of the azimuthal forces:
\beq
2\rho R |N| \Omega \xi_R =  -im\delta P
\eeq
With $k_r=k_\theta = 0$, there are neither components of the Coriolis
force in the $\bb{e_\theta}$ or $\bb{e_r}$ directions, nor are there
unbalanced pressure gradient forces.  The convective rolls occur preferentially
in planforms of latitudinal arcs: the familiar ``banana cells'' often seen in
laboratory experiments and similuations (e.g., Hart et al. 1986).
Rotational forces have no effect on these rapidly growing linear disturbances,
and therefore no effect on the convective stability criterion.  If the squared
\BV frequency is negative, no amount of (uniform) rotation can stabilize 
modes with vanishing poloidal wavenumber components.

\subsubsection{Differential rotation: leading order effects}

The full problem of the evolution of three-dimensional disturbances
in a two-dimensional background medium is a matter of some complexity,
which we defer to the next section.  But for applications to the SCZ,
$|r\del\ln\Omega|\sim 0.1$, and it is appropriate to use this as
a small parameter as a means to calculate and understand
the leading order effects.  
If we take $k_R= k_z=0$ as the zeroth order solution, then the
polodial wavevector $\bb{k_p}$ is (to all higher orders, in fact):
\beq\label{pee64}
\bb{k_p} = - mt\del\Omega.
\eeq
Expanding equations (\ref{ddotxir}) and (\ref{ddotxiz}) to linear
order in the $\Omega$ gradients, we obtain two very simple equations:
\beq\label{jj1}
\ddot{\xi}_R - R{\dd\Omega^2\over \dd z} t  \dot{\xi}_z -{1\over\gamma
\rho} {\dd P\over \dd R}\, (\bb{\xi\cdot\nabla})\sigma = 0.
\eeq
\beq\label{jj2}
\ddot{\xi}_z + R{\dd\Omega^2\over \dd z} t \dot{\xi}_R -{1\over\gamma
\rho} {\dd P\over \dd z}\, (\bb{\xi\cdot\nabla})\sigma = 0.
\eeq
To derive equations (\ref{jj1}) and (\ref{jj2}), note that
terms linear in $\dd\Omega^2/\dd z$ come from the third and fourth
terms in equation (\ref{ddotxir}), and from the third term in
equation (\ref{ddotxiz}).  In the former case, there is a cancellation
of the $\dot{\xi}_R$ terms, leaving a lone
contribution from $\dot{\xi}_z$.  In the latter case, only the $\dot{\xi}_R$
contribution is linear in the $\Omega$ gradient.  In the end,
only one additional term arises in each of equations (\ref{jj1})
and (\ref{jj2}) from the differential
rotation, and in each case only the axial gradient of $\Omega^2$
enters: if the rotation is constant on clylinders ($\Omega=\Omega(R)$), 
there is {\em no} leading order (linear in the $\Omega$ gradient) correction
to the convective displacements.  In other words, {\em baroclinic vorticity 
must be present in the background to obtain a deviation from radial motion
at linear order in the differential rotation parameter.}  

It is instructive to write (\ref{jj1}) and (\ref{jj2}) in terms of
$\xi_r$ and $\xi_\theta$, the (spherical) radial and colatitudinal
displacements.  
If $P\simeq P(r)$, then to first order, (\ref{jj1})
and (\ref{jj2}) combine to give
\beq\label{jjrp}
\ddot{\xi}_r +  R{\dd\Omega^2\over \dd z} t \dot{\xi}_\theta-{1\over\gamma}
{\dd P\over \dd r} (\bb{\xi\cdot\nabla})\sigma= 0,
\eeq
\beq\label{jjrpp}
\ddot{\xi}_\theta -  R{\dd\Omega^2\over \dd z} t \dot{\xi}_r=0,
\eeq
Equation (\ref{jjrpp}) shows that the amplitude of $\xi_\theta$ will be
linear in $\dd\Omega^2/\dd z$. This means that in equation (\ref{jjrp}),
all of the $\xi_\theta$ terms will be {\em second order} in the
$\Omega^2$ gradient.  Since we are working to linear order
in the angular velocity gradient, our equations become 
\beq\label{jjr}
\ddot{\xi}_r +N^2\xi_r = 0,
\eeq
\beq\label{jjth}
\ddot{\xi}_\theta= {g\over r\gamma}{\dd\sigma\over \dd\theta}t \dot{\xi}_r
\eeq
where in equation (\ref{jjth}), we have used thermal wind balance (\ref{vor2})  
to substitute for $R\dd\Omega^2/\dd z$.   
To this order, there is no change in the behavior of the radial component of
the displacement, while a poleward-increasing entropy profile
produces poleward deflections of a radially outward moving
convective displacement (i.e., one that bears excess entropy).
With $|N|t$ of order unity, the 
angular deflections of convective displacements from linear theory
are very similar in magnitude to the observed departures of the 
iso-$\Omega$ surfaces from constant $\theta$ cones in the bulk
of the SCZ.  This, we suggest, is no coincidence: convection
and its hallmark of constant residual entropy
are both intimately associated with constant $\Omega$ surfaces in 
the bulk of the SCZ (BLW), 
and a poleward deflection of the fluid elements
is unavoidable when $\dd\Omega/\dd z <0$.  
The interesting point, as we have earlier noted,
is that it seems there must be an external source of vorticity
in place to drive the deflections.

\section {Numerical solutions}              

\subsection{Representative parameters}

We next consider exact numerical solutions of equations (\ref{ddotxir})
and (\ref{ddotxiz}).
There are four important solar model parameters that 
need to be fixed: the two components of the $\Omega$ gradient,
and the two components of the $\sigma$ (entropy) gradient.  
The $\Omega$
gradient components at a particular location may be read directly from the 
helioseismology data.  The term $\dd\sigma/\dd\theta$ then follows
from the assumption of thermal wind balance.    
Finally, $\dd\sigma/\dd r$ is taken from a published benchmark
solar model (Stix 2004).  Typical values for the $\Omega$ gradient
at midlatitudes near $r=0.85 R_\odot$ are (Christensen-Dalsgaard
\& Thompson 2007):
\beq\label{refgg}
{\dd \ln\Omega\over \dd\ln R} \simeq  0.24\qquad
{\dd \ln\Omega\over \dd\ln z} \simeq  - 0.12
\eeq
Thermal wind balance (i.e., vorticity conservation) then gives 
\beq
{\dd\sigma\over \dd\theta} \simeq  -4\times 10^{-6}
\eeq
while a standard mixing length model (Stix 2004) gives:
\beq\label{stx1}
{\dd\sigma\over \dd\ln r} \simeq  -2.3 \times 10^{-5}
\eeq
Thus, for $\Omega=400$nHz,
\beq\label{refggg}
4\Omega^2 \simeq  2.5\times 10^{-11} {\rm s}^{-2}, 
\qquad {g\over \gamma r}{d\sigma_r\over
d\ln r} \simeq -8.7\times 10^{-12} {\rm s}^{-2}
\eeq
(In the context of our model, once the gradients of $\Omega^2$ have
been taken directly from the helioseismology results and the $\theta$
derivative of $\sigma'$ from thermal wind balance, the $r$ derivative
of $\sigma'$ is uniquely determined from requiring counteraligned
gradients of $\Omega^2$ and $\sigma'$.  Then, equation (\ref{stx1}) is
understood as the radial gradient of $\sigma_r$.)  For the particular
values in equations (\ref{refgg})--(\ref{refggg}), at latitudes less
than $54^\circ$, the axisymmetric displacement solutions of equation
(\ref{asymp}) are stable.

\subsection{Results}
We integrate equations (\ref{ddotxir}) and (\ref{ddotxiz}) for a variety
of different initial velocities and wavenumbers.  Our strategy is to
choose initial wavenumbers lying in the plane tangent to $\bb{e_r}$, with
a random initial velocity direction (perpendicular to the wavenumber),
and an initial displacement of zero.  The ensuing trajectories are
then followed.

When the angular velocity is uniform and free of shear, the results are
very simple: all trajectories rapidly become radial, regardless of their
initial condition.   In agreement with our analytic treatment, 
after a brief initial transient, there is
no equatorial or poleward deflection, even with the Coriolis force.
On the other hand, when an angular velocity profile is used that
has been modeled with the helioseismology data, there are $\sim$10\%
poleward deflections on time scales of a month or two, just as equations
(\ref{jjr}) and (\ref{jjth}) would predict.


\begin{figure}[h]
\centering
\caption{Meridional slices of the Sun ($r=0.8R_\odot$) at four different times
(in units of $2\pi/\Omega$) showing the relative orientation
of the radial direction (dotted black line), entropy flux (solid lines), and 
isorotation surfaces (dotted red lines).  Everywhere but at the highest
latitudes, the entropy flux aligns itself with the isorotation surfaces within
a typical mixing time.
See text for further details.}
\includegraphics[width=10cm]{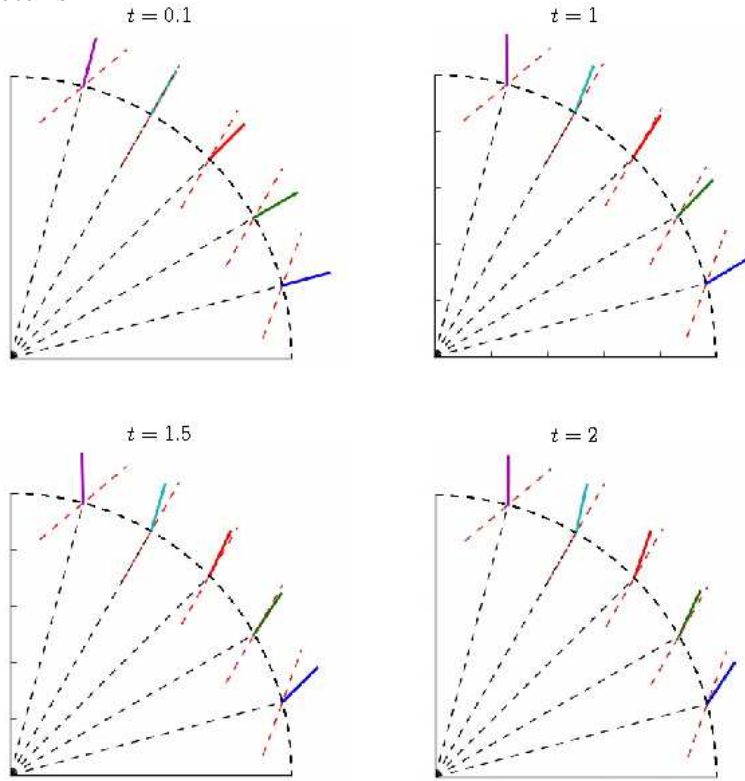}
\label{sunslice}
\end{figure}

Figure (1) shows this effect clearly.  We work at $r=0.8 R_\odot$, at latitudes
$15^\circ$, $30^\circ$, $45^\circ$, and $60^\circ$.  At each latitude, the surface
of the local isorotation curve is shown, edge-on in a meridional plane,
by a dotted red line.  The local radial direction is shown by the black dotted
line.  The colored solid line at each latitude is the direction of the local
{\em Eulerian} perturbation velocity,
projected in the meridional plane.  Four different times are shown.
In each case, the trajectory passes through the local isorotational
surface on a time scale of a few $e$-foldings, i.e. something close to
a mixing time.  Near the pole---above $\sim 60^\circ$---the trajectories
begin and remain northward of the local iosortation surface, becoming more
so with time.  Here, the problem becomes close to one-dimensional, with
all flow quantities predominantly a function of $z$.  Our fundamental
assumption that there is a functional relationship between angular
velocity and residual entropy is then a matter of mathematical symmetry,
not dynamics.

The torques required to maintain the fluid elements in or near surfaces of constant
angular velocity are provided by azimuthal pressure gradients, as previously noted.  
{\em In no
sense is the angular momentum of an individual fluid element conserved.}  Consider
however $\delta v_\phi$,
the Eulerian perturbation of the angular velocity:
\beq
\delta v_\phi = {\dot\xi}_\phi -(R\bb{\xi\cdot\nabla})\Omega =
-R(k_r\dot\xi_r+k_\theta \dot\xi_\theta)/m,
\eeq
where we have used equation (\ref{xiphidot}) in the last equality, and
switched to spherical coordinates.  The wavenumber component $k_r$ is very
small in the bulk of the SCZ since it is proportional to $\dd\Omega/\dd r$
and $\Omega\simeq\Omega(\theta)$.   The term $k_\theta\dot\xi_\theta$
term is quadratic in the $\Omega$ gradients, thus also a very small
quantity.  Therefore, there is very little angular momentum transport
propagated by the correlated fluctuation tensor component $\dot\xi_r\delta
v_\phi$.  Indeed, noting that if equation (\ref{pee64}) holds for
the poloidal wavenumber components, then
\beq
\delta v_\phi =-R(k_r\dot\xi_r+k_\theta \dot\xi_\theta)/m=Rt\bb{\dot\xi}
\bcdot\del\Omega,
\eeq
which for a near vanishing $\delta v_\phi$
is a self-consistent indication that displacements do not stray from
constant $\Omega$ surfaces.
Angular momentum could, in principle, still be directly advected
by a nonvanishing poloidal mass flux, if one is present.

\subsection{Linear convection theory: a summary}

The linear behaviour of perturbations in a two-dimensional, stratified,
differentially rotating medium is very different according to whether
or not $\Omega$ is dependent upon $z$.  If $\Omega$ depends only upon
$R$, isobaric and isochoric surfaces coincide, there is no source
of baroclinic vorticity, and the dominant convective displacements are radial.
There is neither poleward nor equatorial bias in the heat transport.
If $\Omega$ depends upon $z$, isobaric and isochoric surfaces do not
coincide, an explicit vorticity source is implicated, and outward moving, hotter
convective displacements deflect north or south depending upon whether
$\Omega$ respectively decreases or increases poleward.

The helioseismology data in the bulk of the SCZ are in accord with this
description, with $\dd \Omega^2/\dd z <0$ and slightly upward cants to the
constant $\Omega$ surfaces.  The question remains, however, of what the
cause of this gradient is.  Indeed, it is a classical case of begging
the question: the fundamental angular entropy gradient, putatively arising
from the interaction of rotation and convection, itself requires the
prexistence of an axial $\Omega$ gradient---the very gradient
the angular entropy gradient is supposed to be explaining!

In the next section, we suggest a resolution to this problem.

\section{Vorticity generation}
\subsection {Centrifugal distortion of stratified surfaces}

It has long been known that a uniformly rotating star
cannot simultaneously be both in radiative
and hydrostatic equilibrium (e.g. Schwarzschild 1958).  
The difficulty is that the rotation induces centrifugal distortions of the
isothermal surfaces---polar flattening and equatorial bulging---which are
incompatible with a vanishing radiative flux divergence.  The dimensionless
number that sets the scale for these distortions is the centrifugal parameter,
which for a (fictional) uniformly rotating Sun takes the value:
\beq
\epsilon_0 = {R_\odot^3 \Omega^2\over GM_\odot} \sim 1.6\times 10^{-5}.
\eeq
(We have used $\Omega= 2.5\times 10^{-6}$.)  This is a very small
number.   But we are interested in entropy
gradients of order 
\beq
{\dd\sigma\over\dd\theta}\sim 10^{-6},
\eeq
and it therefore behooves us
to take note of centrifugally-induced angular gradients.  

Normally, a tiny amount of meridional circulation is enough to offset the
unbalanced radiative heating.  Consider, however, conditions at the 
outer edge of the radiative zone.  The entropy equation
is
\beq
P \left[ {\dd \sigma\over \dd t} +(\vv\bcdot\del)\sigma
\right]= -(\gamma-1)\del\bcdot\bb{F},
\eeq
where $\bb{F}$ is the radiative flux.  If the divergence term on the
right does not vanish at the radius where the entropy gradient does,
there is no simple balance of the right and left sides of the
equation.  A balance could in principle be restored if a turbulent
entropy flux divergence were present, and this may indeed be representative of
current conditions, but it is revealing to follow the breakdown of our
simple uniform rotation model.  Uneven heating would alter the flux until
its divergence is minmized.  (The time scale for this is simply the
Kelvin-Helmholtz time; $\epsilon_0$ is not involved.)  This altered flux
divergence will generally be incompatible with coincidence of isobaric
and isochoric surfaces, which uniform rotation demands.  We are thus led
to the generation of differential rotation with an axial component of the
angular velocity gradient (baroclinicity) to maintain thermal equilibrium.
Even in a more complex setting with a thermal entropy flux divergence
present, the base of the convective zone will follow the 
the radiation flux divergence and almost certainly be more
spherical than the equipotential surfaces.  Once again baroclinic
structure will be generated\footnote{We acknowledge the referee M. Miesch
for emphasizing this point.}.

The helioseismology data show unambiguously that the radius at which
the entropy gradient vanishes is symmetrically placed within a narrow
band of pronounced differential rotation: the tachocline.  Moving from
this radius downward into the radiative zone, uniform rotation takes
over as the entropy gradient rises rapidly and meridional circulation
is established.  Moving upward into the convective zone, turbulent
mixing quickly develops and there is an abrupt change of the character
of the differential rotation: convective mixing leads at once to the
coincidence of two important classes of surface, but not a coincidence
that is compatible with uniform rotation.  Rather, it is isorotational
and residual entropy surfaces that now coincide.  (Note the hidden
but important role of nonaxisymmetry, as the convective modes are
large $m$ disturbances.)  Turbulent convection,
in this picture, does not generate ``from scratch'' the differential
rotation in which it operates.  Instead it reacts to, and reinforces,
the angular velocity gradient bequeathed to it from the surface of
vanishing entropy gradient.  We suggest that this is a key ingredient
to the organizational scheme of differential rotation in the Sun.

\subsection{Octopolar structure}     

The centrifugal distortion of equipotential surfaces engendered by uniform
rotation leaves temperature $T$, pressure $P$ and density $\rho$ functions
of $r[1+\epsilon(r)P_2(\cos\theta)]$, where $P_2$ is the usual
Legendre polynomial of second order, and $\epsilon$ is a function only
of radius $r$ (of order $\epsilon_0$).
The function $\epsilon$ is determined by the solving the Poisson equation
with appropriate boundary conditions (Schwarzschild 1958).   
At first glance it might be thought that the effect of vorticity
generation and its differing
iso-surfaces would be to create a distinct
$\epsilon$ for each structural variable: $\epsilon_\rho(r)$, $\epsilon_T(r)$, etc.  
In fact, when vorticity is generated
it is impossible to satisfy basic rotational equilibrium without
a $P_4(\cos\theta)$ dependence in the {\em leading order} nonspherical structure
of these variables.  

The equation of rotational equilibrium is
\beq
R\Omega^2 \bb{e_R} = {1\over\rho}\del P + \del\Phi
\eeq
where $\Omega$ depends upon $r$ and $\theta$.  From the symmetry of our problem
we would expect $\Omega^2$ to be of the form
\beq
\Omega^2 = \Omega_0(r)^2 + q_2(r) P_2(\cos\theta)
+q_4(r) P_4(\cos\theta) + ...
\eeq
where the $q_{2i}$ are functions of $r$ only.
This is not necessarily an expansion in a small parameter,
but for the region of interest the first two term provide 
a very good approximation,
with errors in $\Omega^2-\Omega_0^2$ about $10$\% very near the equator and
less elsewhere\footnote{
Numerical fits generally are quoted for $\Omega$ rather than $\Omega^2$;
the latter turns out to have a simpler expansion near the tachocline boundary.  
See Gough [2007] for
convenient parameterizations of $\Omega$.}.
The average of $\Omega^2$ is the average of $\Omega_0^2$ and will be denoted
$\overline{\Omega^2}$.  Then, we may write the force balance equation as
\beq\label{bar1}
R(\Omega^2- \overline{\Omega^2}) \bb{e_R} = {1\over\rho}\del P + \del\Phi'
\eeq
where
\beq
\Phi'= \Phi -{R^2\over2}\overline{\Omega^2}
\eeq
Taking the divergence of equation (\ref{bar1}),
\beq
{1\over R}{\dd\ \over \dd R}\left[R^2(\Omega^2- \overline{\Omega^2})\right]
=\del\bcdot\left({1\over\rho}\del P\right) + 4\pi G\rho - 2\overline{\Omega^2}
\eeq
or
\beq
{1\over r}\left( {\dd \over \dd r} + {\cot\theta\over r}{\dd\ \over \dd\theta}
\right)\left[r^2\sin^2\theta(\Omega^2- \overline{\Omega^2})\right]
=\del\bcdot\left({1\over\rho}\del P\right) + 4\pi G\rho - 2\overline{\Omega^2}
\eeq
From the form of this equation, it is evident that if $\Omega^2$ has terms
through order $P_l(\cos\theta)$ in its angular expansion, 
then $\rho$ and $P$ will in general
have angular structure through $P_{l+2}$ in the first order linear term of a
small $\epsilon_0$ expansion.  Therefore, uniform rotation results in
quadrupolar deformation of the iso-surfaces of all structural variables,
whereas a simple $\sin^2\theta$ latitudinal dependence of $\Omega^2$
results in {\em octopolar} structure.

Consider a scenario in which the Sun is rotating uniformly.  There are
$P_2$ distortions of iso-surfaces in the radiative zone, and no polar
deflections of warm fluid elements in the convective zone.  Baroclinic
vorticity would then be produced at the radius of vanishing entropy
gradient.  Were only $P_2$ structure to remain, the only self-consistent
solution for the angular velocity would be $\Omega=\Omega(r)$.  However,
$\dd\Omega/\dd z <0$ would now also be present, causing first order
poleward deflections of warm convective elements.  

These results are very suggestive, and offer at least a heuristic
approach to understanding the Sun's poleward decreasing angular velocity
profile.  If $\dd \Omega^2/\dd z<0$ is maintained through the tachocline
into the bulk of the SCZ, in which convection largely eliminates the
$r$ gradients of $\Omega$ and $\sigma'$, then $\dd\Omega^2/\dd\theta$
must be positive.  To justify these assumptions in detail, however, one
would need to know how angular momentum is transported and deposited by
secondary flows (Meisch et al. 2012), and what is relative importance
of vorticity forcing by nonconservative forces versus inertial vorticity
conservation.  These considerations, whose detailed origin lies outside
the the scope of the current work, regulate the the $\theta$ dependence
of the pressure, density, entropy, and ultimately the angular velocity.

\subsection {Solution for $\Omega^2$}

The above considerations suggest that we regard the $\rho$ (for example)
as a function of the quantity
\beq
 r_\rho \equiv r[1+\epsilon_\rho(r)f_\rho (\cos^2\theta)],
\eeq
and similarly for $P$ and $T$.  
The $f_i(\cos^2\theta)$ functions are linear combinations of $P_2$ and
$P_4$, in general distinct for $i=\rho, P, T$.   (Recall that in the
numerical simulations described by Miesch et al. (2006), $P_l$ angular
structure in the {\em entropy} was included as a boundary condition at the
base of the convection zone.  By far the best solar fit included $P_2$
and $P_4$ terms in the angular structure of the entropy.)  Expanding $r_\rho$, 
\beq
\rho(r_\rho) = \rho_0(r) +\delta \rho(r) +  \epsilon_\rho(r){d\rho_0\over d\ln r}
f_\rho (\cos^2\theta) +...
\eeq
where $\rho_0$ is the nonrotating solution and $\delta\rho = \rho(r)-\rho_0(r)$.
To leading order,
\beq
{\dd\rho\over \dd r} = {d\rho_0\over dr}, \qquad {\dd\rho\over \dd\theta} = 
-2\cos\theta\sin\theta \, \epsilon_\rho(r)\ {d\rho_0\over d\ln r}\,
f'_\rho (\cos^2\theta),
\eeq
with similar results for $P$ and $T$.  The notation $f'$ denotes a derivative
with respect to $\cos^2\theta$.  The $f'$ functions are thus
a linear superposition of $\cos^2\theta$ (or if more convenient $\sin^2\theta$)
plus a constant term.

With this development, the vorticity equation becomes
\beq\label{exacto}
{\dd\Omega^2\over \dd z}  = 
2\cos\theta\> (\epsilon_P f'_P -\epsilon_\rho f'_\rho){1\over
\rho_0 r} {dP_0\over dr} {d\ln \rho_0\over dr} .
\eeq
The demands of
radiative equilibrium will require isothermal surfaces to
be more spherical (less distorted) than are isochoric surfaces, whose
centrifugal deformations are less critical to the heat flux.
The pressure, being a product of $\rho$ and $T$, would thus also
be closer to spherically stratified than $\rho$ itself.  
A simple mathematical approximation that allows progress and
respects the observational
fundamentals is to assume that the $f$ stratification
functions are the same for
each variable, but that the $\epsilon$ functions differ in magnitude
from one variable to the next: $\epsilon_\rho>\epsilon_P>\epsilon_T$.
We will also ignore the spatial variation of the $\epsilon$ within
the tachocline boundary layer.  

These considerations lead to a vorticity equation of the form
\beq\label{clunk}
{\dd\Omega^2\over \dd z} = {A\over r^4} \cos\theta(\sin^2\theta - \alpha)
\eeq
where $A$ (of order $\epsilon_0 GM_\odot$) and $\alpha$
(of order unity) are positive constants.  We have assumed $1/r^2$ gravitational
field and a $1/r$ dependence for $d\ln\rho/dr$.  Signs have been chosen
so that $\dd\Omega^2/\dd z$ is negative at non-equatorial latitudes, as
indicated by the helioseismology data.    Another way to write this
equation is
\beq\label{clunk1}
{{\cal D}\Omega^2\over {\cal D} r} = {A\over r^4} (\sin^2\theta- \alpha)
\eeq
where the characteristic derivative ${\cal D}/{\cal D} r$ is taken along the 
path
\beq\label{clunk2}
r^2\sin^2\theta = {\rm Constant\ } \equiv r_0^2\sin^2\theta_0
\eeq
In this form, the equation is close to the tachocline equation (19) of
BLW, which is ``correct'' (in the sense that it
agrees extremely well with the data), but derived from a completely different
point of view.  The sole difference in the two formalisms 
is that the ${\cal D}/{\cal D}r$ derivative of BLW is taken along the path 
\beq\label{clunk3}
r^2\sin^2\theta =  r_0^2\sin^2\theta_0 +\beta r_0^2\left(1 -{r_0\over r}\right)
\eeq
where $\beta$ is a number of order unity.  This is the characteristic path
associated with the SCZ isorotation contours.  

How closely do the two approaches agree?  If $\Omega^2$ is a given function
$\Omega_0^2(r_0^2\sin^2\theta_0)$ at radius $r=r_0$, then
the solution of (\ref{clunk1}) and (\ref{clunk2}) is 
\beq
\Omega^2 = \Omega_0^2(r_0^2\sin^2\theta_0) + A \int^r_{r_0}
\left( {r_0^2\sin^2\theta_0\over r^6} - {\alpha\over r^4}\right)\, dr
\eeq
or
\beq
\Omega^2 = \Omega_0^2(r_0^2\sin^2\theta_0)+{A\sin^2\theta_0\over 5 r_0^3}
\left( 1 - {r_0^5\over r^5}\right) -{A\alpha\over 3 r_0^3}
\left(1- {r_0^3\over r^3}\right)
\eeq
Next assume that 
\beq
\Omega^2_0 \equiv \Omega_1^2 +\Omega_2^2\sin^2\theta_0
\eeq
where $\Omega_1^2$ and $\Omega_2^2$ are positive constants.
(Recall that this is an excellent approximation for the boundary of the 
tachocline.)
Using equation (\ref{clunk2}) and solving for the isorotation curves
leads to
\beq\label{96er}
\sin^2\theta = {r_0^2\over r^2} \left( \Delta + C\alpha F_3\over 1 + 
CF_5\right)
\eeq
where
\beq
\Delta = (\Omega^2-\Omega_1^2)/ \Omega_2^2, \quad C = A/(\Omega_2^2 r_0^3),
\quad F_j = (1/j)[1-(r_0/r)^j]
\eeq
Note that $\Delta$ is numerically the same as $\sin^2\theta_0$;
$C$ is a number of order $\epsilon_0 \times 1/\epsilon_0\sim$ unity.

\begin{figure} [h]
\begin{center}
\epsfig{file=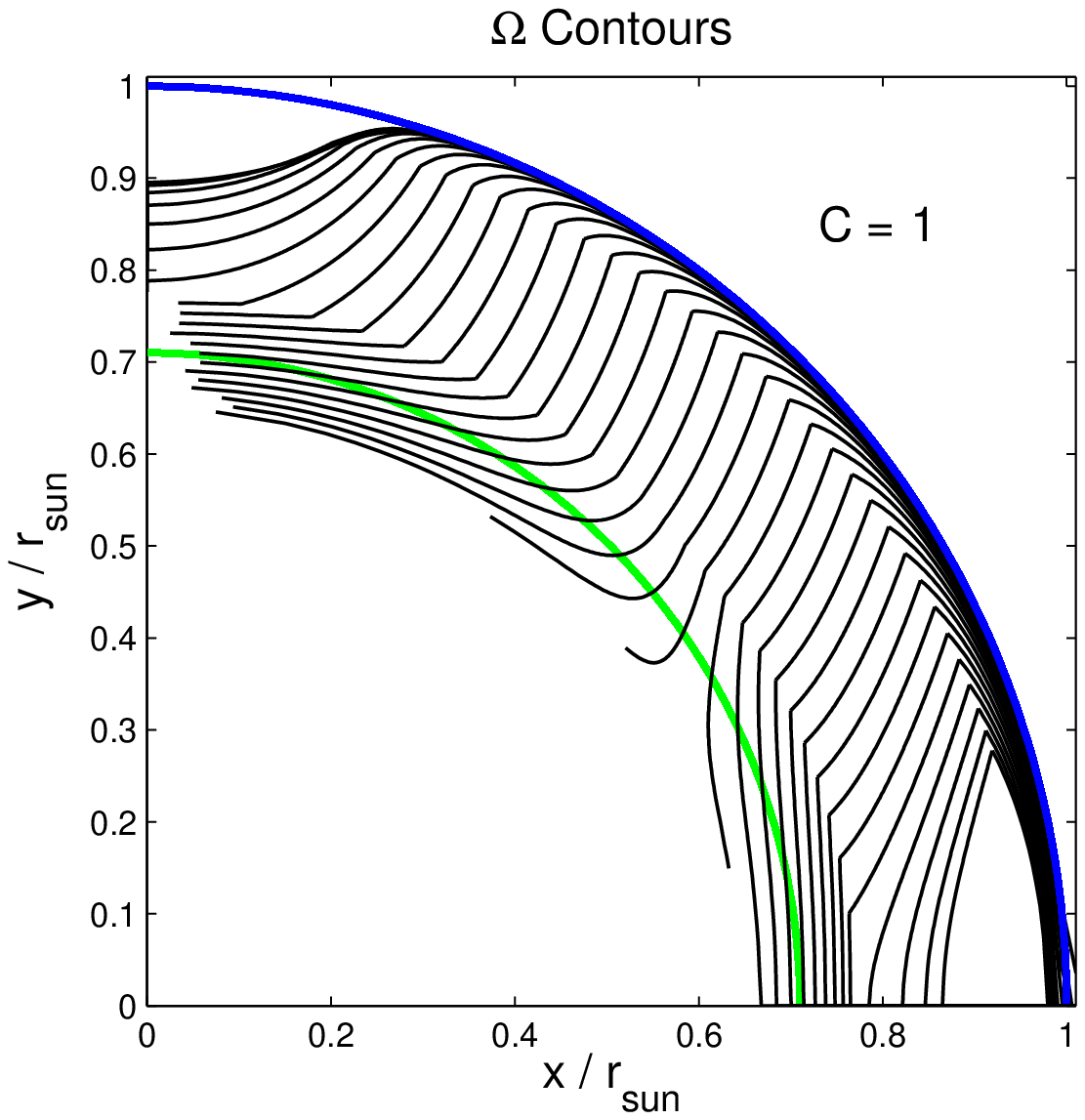, width=9.0 cm}
\epsfig{file=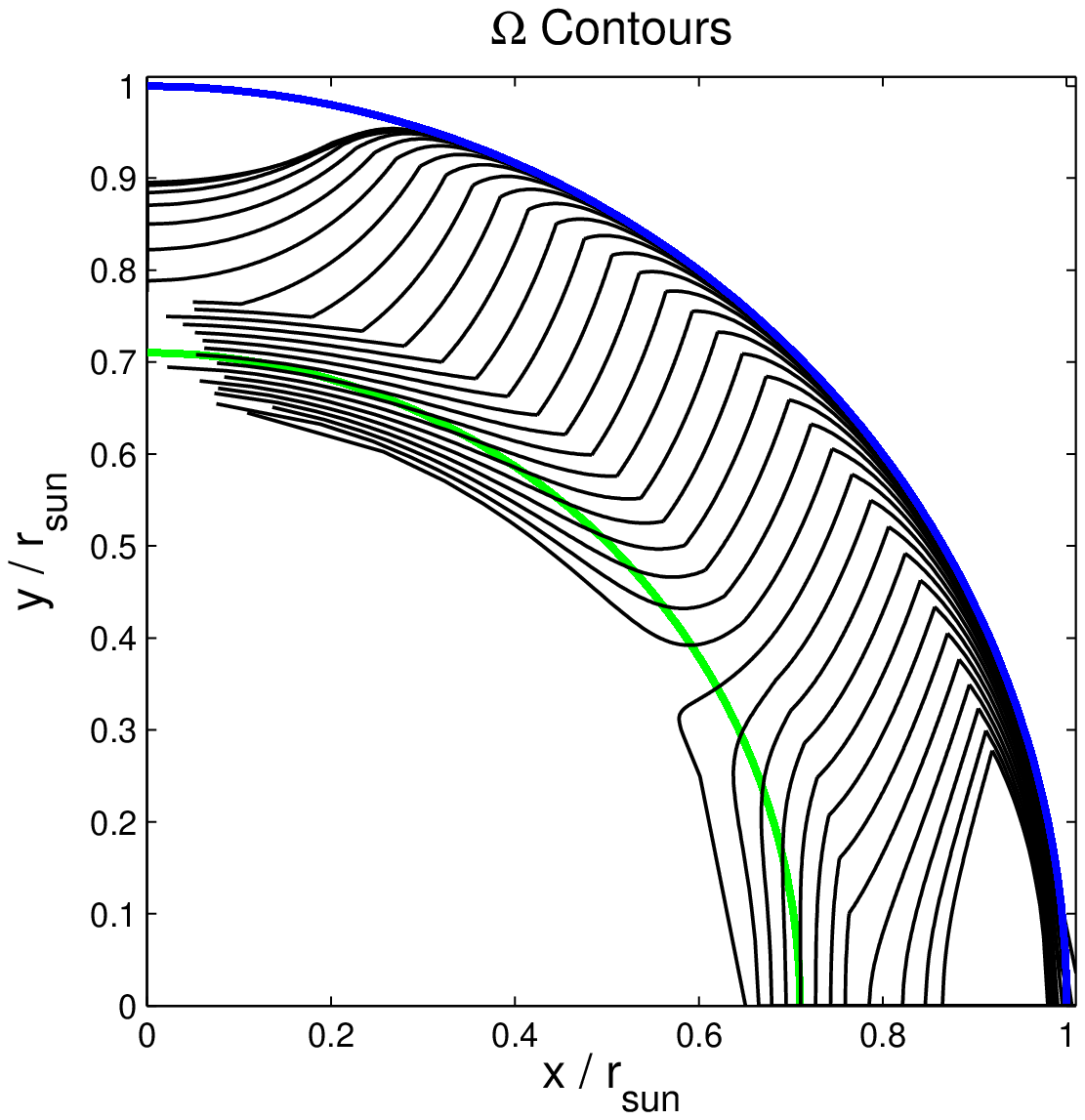, width=9.0 cm}
\epsfig{file=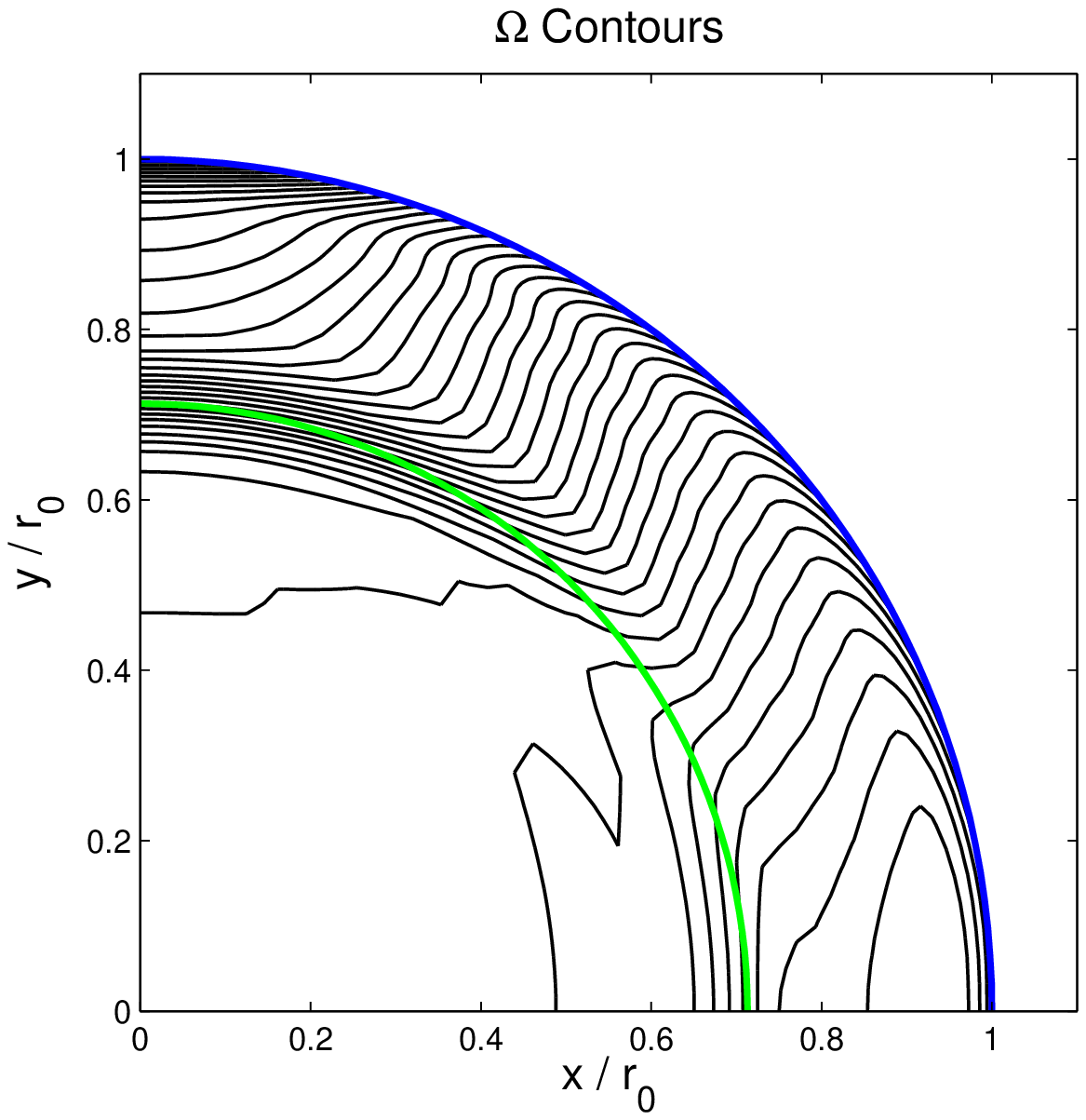, width=9.0 cm}
\caption {{\em Top:} The $C=1$, $\alpha=0.8$ tachocline solution of equation (\ref{clunk1})
along the trajectory characteristics of (\ref{clunk2}).
Tachocline solutions are joined, at $r=0.77R_\odot$, to the best fit solution of BLW 
for the SCZ and outer layers.  Formal tachocline location is indicated
by green line.  {\em Middle:} The best fit global solution of BLW, in good agreement
with observations. 
In this case, equation (\ref{clunk1}) is solved along characteristics given by
(\ref{clunk3}).  {\em Bottom:} GONG data, courtesy R. Howe.}
\end{center}
\end{figure}

As can be seen in figure (2), which shows our solution (\ref{96er})
next to the precise fit of BLW, the essential features of our result
are basically correct.  We have embedded the tachocline solution inside
the same convective envelope solution in both cases.  The principal
area of disagreement is the bifurcation zone, on one side of which the
isorotation contours break toward the pole and on the other side toward
the equator.  This is clearly the region where a crude approximation of
the right side of equation (\ref{exacto}) is likely to be most inaccurate.
Our approach to the tachocline structure makes it more clear why the BLW
equation works so well.  An uncertain approximation used by BLW was the extension
of the same simple functional dependence between $\sigma'$ and $\Omega^2$
into the tachocline.  But in the region of interest near the bifurcation point, the
tachocline solution really does become a smooth extension of the SCZ (and the 
tachocline belies its name).   Away from the bifurcation zone, the
functional dependence hardly matters, as it affects only the subdominant
$\dd \Omega^2/\dd\theta$ term in the governing equation.  

The main point that one should take away from this exercise is that the
combination of direct vorticity forcing together with a proper reckoning
of the octopolar distortions of the stuctural variables $P$ and $\rho$
caused by the centrifugal forces of even the simplest latitude-dependence
of the angular velocity seem to be important components of the rotational
profile of the tachocline.

\section {Conclusions}

The current work combines two very different types of calculation,
linking the linear dynamics of the convective zone to the origins
of solar differential rotation and vorticity.  We begin with the second
part first.

We have argued that the centrifugal distortion of equipotential surfaces
combined with demands of thermal equilibrium requires the cleaving of
isobaric and isochoric surfaces, and is likely to be the underlying cause
of the Sun's differential rotation.  Current numerical simulations are
not designed to capture a process in which an order $\epsilon_0\sim
10^{-5}$ radiative effect is turned into relative angular velocity
gradients of order $0.1$.  It may well be possible, however, to devise
other computational schemes tailored to working with this point-of-view.

An important technical point is that the tachocline pressure and
density (and therefore entropy) must have both $P_2(\cos\theta)$ and
$P_4(\cos\theta)$ angular structure.   Although this in itself is not
a particularly new result, neither has it been widely appreciated,
and we have exploited it in a rather novel manner.  With $P_4$ and
$P_2$ structure in place, not only may one understand the simple
form of the vorticity equation in the tachocline, with 
reasonable approximations one may explicitly solve the equation.
An important parameter of this equation, namely the precise angle at which
$\dd\Omega^2/\dd r$ changes sign, is probably determined by minimizing
the torque on the radiative interior.  Finally, it is interesting to note
that Roxburgh (2001) calculated the Sun's multipole moments using models
based on helioseismic inversions.  He found that the size of the octopole
term $J_4$ was comparable to the {\em change} in the quadrupole term $J_2$
when differential rotation was included.  This is what one would expect
if differential rotation were {\em modifying} $P_2$ and {\em creating}
$P_4$ angular structures.

In the first part of this paper, we have carried out a very general
Lagrangian linear calculation of the fluid displacements in an arbitrary,
magnetized, two-dimensional background, stratified in both $R$ and $z$
(or $r$ and $\theta$).  Applied to conditions appropriate to the SCZ,
in the absence of a background $\dd\Omega/\dd z$ gradient, we find no
poleward deflection of hot convective fluid elements.  Such an effect is
sometimes invoked to produce an angular entropy gradient which would in
turn lead to differential rotation via thermal wind balance.  Although
there is nothing in our calculations that would prohibit the emergence
of the required gradients in $\Omega$ and $z$ at nonlinear order 
in a turbulent fluid, it is some significance that the dominant leading
order linear response is completely different depending upon whether
$\dd\Omega/\dd z$ is present or absent.  When a finite $\dd\Omega/\dd z$
is {\em a priori} present, there are reinforcing convective deflections:
a negative axial $\Omega$ gradient, in particular, engenders poleward
deflections of warmer fluid elements, which via thermal wind balance
strengthen and maintain this same $\Omega$ gradient.  The presence
or absence of background baroclinic vorticity is thus mirrored in the
leading behavior of convective displacements.  

The actual generation
of baroclinic vorticity may well lie outside the realm of convection
dynamics.  We propose that a vorticity source will inevitably appear at
the location of a vanishing entropy gradient.  
In general, because of the effects centrifugal flattening, the constraints
of thermal and dynamical equilibrium will force
different iso-surfaces for density, temperature, and pressure.  This must
lead to an axial angular velocity gradient.  By making some simplifying
but plausible approximations, one may calculate a time-steady solution 
of the vorticity
equation for the angular velocity.  This explicit solution 
yields a tachocline structure that certainly resembles the
observations, and bears comparison with an earlier, more accurate, but
also more phenomenological, calculation (BLW).  With the onset of fully
developed convection, surfaces of constant angular velocity and residual
entropy coincide, and the character of the isorotation contours takes on
the classical conical form well-known from helioseismology.

The strength of this view of the origin of solar differential is
that it emerges from just a few rather simple and largely inevitable
processes: the demands of thermal energy balance in a rotating
system (which creates a baroclinic axial gradient of $\Omega$ at the
radiative/convective boundary), the kinematics of shear (which, via embedded
wavenumbers, incorporates $\Omega$ gradients into the response of the fluid
displacements), and the linear dynamics of convection (which causes poleward
displacements of entropy bearing fluid elements).                              
While direct numerical simulation of this scenario is likely to be very 
challenging, it may well be possible to design a test-of-principle proxy system.

\section*{Acknowledgments}
Much of this work was carried out while SB was a Paczynski Visitor
and ES a visiting student researcher in the Department of Astronomy at
Princeton University.  We are grateful to Profs.\ D.\ Spergel and
J. Stone for generous support during this visit.  This work has also
benefitted by grants from the Institut universitaire de France and
the Conseil R\'egional de l'Ile de France, and especially benefitted
from a very constructive critique by our referee M.\ Miesch.  It is a
pleasure to acknowledge helpful conversations with H.\ Latter, P.\ Lesaffre,
E.\ Quataert, J.\ Stone and N.\ Weiss, and to thank G.\ Mirou for spotting
a technical error in an earlier draft of this work.

\section*{Appendix}
To recover the axisymmetric limit of equations (\ref{ddotxir}) and
(\ref{ddotxiz}), we begin with their equivalent forms (\ref{I}) and
(\ref{II}):
\beq\label{Ibis}
\ddot\xi_R +(\kva)^2\xi_R -{k_R\over k_z}\left(\ddot\xi_z +(\kva)^2\xi_z
\right)
+{2\Omega R\over m}\left( k_R\dot\xi_R+k_z\dot\xi_z\right)
+{ {\cal D}P\over\rho\gamma} (\bb{\xi}\bcdot\del) \sigma
=0,
\eeq
$$
{Rk_R\over m} \left[\ddot\xi_R +(\kva)^2\xi_R\right]
+\left( {Rk_z\over m}+{m\over Rk_z}\right)
\left[\ddot\xi_z +(\kva)^2\xi_z\right]
\qquad\ \ \
\qquad\ \ \
\qquad\ \ \
\qquad\ \ \
\qquad\ \ \
\qquad\ \ \
$$
\beq\label{IIbis}
\qquad
\qquad
\qquad
\qquad
\qquad
\qquad
\qquad
-2\bb{\dot\xi\cdot\nabla}(R\Omega) +
{m\over Rk_z\rho\gamma}
{\dd P\over \dd z}
{ (\bb{\xi}\bcdot\del)} \sigma
=0,
\eeq
In the axisymmetric $m\rightarrow0$ limit, equation (\ref{Ibis}) 
may be written
\beq\label{Aone}
{2\Omega R\over m} \left(k_R\dot\xi_R + k_z\dot\xi_z\right)  + {k^2\over k_z^2}
\left[ \ddot\xi_R +(\kva)^2\xi_R\right]
+{ {\cal D}P\over\rho\gamma} (\bb{\xi}\bcdot\del) \sigma
=0,
\eeq
In what follows, we will also require the twice differentiated form
of this equation,
\beq\label{Atwo}
{2\Omega R\over m} \left(k_R \xi^{iii}_R + k_z \xi^{iii}_z\right)
-2R\bb{\ddot\xi}\bcdot\del\Omega^2 + {k^2\over k_z^2}
\left[\xi^{iv}_R +(\kva)^2\ddot\xi_R\right]
+{ {\cal D}P\over\rho\gamma} (\bb{\ddot\xi}\bcdot\del) \sigma
=0,
\eeq
where the notation $iii$ and $iv$ denotes three and four time
differentiations, respectively.  Only the leading order
terms in a small $m$ expansion have been retained.

Now, in the axisymmetric limit, equation (\ref{IIbis}) becomes
\beq\label{Athree}
{R\over m}\left( k_R\ddot\xi_R + k_z\ddot\xi_z\right)+(\kva)^2
{R\over m}\left( k_R\xi_R + k_z\xi_z\right)- 2 \bb{\dot\xi\cdot\nabla}
(R\Omega)=0
\eeq
Differentiating once,
\beq\label{Aiv}
{R\over m}\left( k_R\xi^{iii}_R + k^{iii}_z\xi_z\right)+(\kva)^2
{R\over m}\left( k_R\dot\xi_R + k_z\dot\xi_z\right)
-R \left[\bb{\ddot\xi} + (\kva)^2\bb{\xi}\right]\bcdot\del\Omega
- 2 \bb{\ddot\xi\cdot\nabla}
(R\Omega)=0.
\eeq
The axisymmetric dispersion relation now follows from substituting
equations (\ref{Aone}) and (\ref{Atwo})
for the $1/m$ terms
into equation (\ref{Aiv}), setting
$\xi_z=-k_R\xi_R/k_z$, and replacing all time derivatives by $-i\omega$.
(The sign is chosen so that a positive wavenumber has a
positive phase velocity.) 
After algebraic simplification, the result is
\beq
{k^2\over k_z^2}\varpi^4 +\varpi^2\left[ {1\over R^3}{\cal D}(R^4\Omega^2)
+{1\over\rho\gamma}({\cal D}P)(  {\cal D} \sigma)\right] -
4\Omega^2(\kva)^2 = 0, \qquad \varpi^2 =\omega^2 -(\kva)^2,
\eeq
where ${\cal D}$ is defined in equation (\ref{Deq}).
This is in precise
agreement with Balbus (1995).


\begin{thebibliography}{99}

\bibitem[\protect\citeauthoryear{Balbus}{1995}]{b95}Balbus, S. A., 1995, ApJ, 453, 380

\bibitem[\protect\citeauthoryear{Balbus et al.}{2009}]{bblw}
Balbus, S. A., Bonart, J., Latter, H. N., Weiss, N. O., 2009, MNRAS, 400, 176 

\bibitem[\protect\citeauthoryear{Balbus \& Hawley}{1994}]{bh94}Balbus, S. A., Hawley, J. F. 1994, MNRAS, 266, 769 

\bibitem[\protect\citeauthoryear{Balbus et al.}{2009}]
{blw}Balbus, S. A., Latter, H., Weiss, N., 2012, MNRAS, 420, 2457 (BLW)     

\bibitem[\protect\citeauthoryear{Christensen-Dalsgaard \& Thompson}{2007}]{ct07}
Christensen-Dalsgaard, J., Thompson, M. J., 2007, in The Solar Tachocline, eds. D. Hughes, R. Rosner, N. Weiss, 
(Cambridge University Press: Cambridge), p. 53

\bibitem[\protect\citeauthoryear{Cowling}{1951}]{c51}Cowling, T. G., 1951, ApJ, 114, 272

\bibitem[\protect\citeauthoryear{Goldreich \& Schubert}{1967}]{gs67}Goldreich, P., Schubert, G., 1967, ApJ,150, 571

\bibitem[\protect\citeauthoryear{Gough}{2007}]{g07}
Gough, D. O., 2007, in The Solar Tachocline, eds. D. Hughes, R. Rosner, N. Weiss 
(Cambridge: Cambridge University Press), p. 3

\bibitem[\protect\citeauthoryear{Gough \& McIntyre}{1998}]{gm98} Gough, D. O., McIntyre, M. E., 1998, Nature 394, 755

\bibitem[\protect\citeauthoryear{Hart et al.}{1986}]{hetal86} Hart J. E., Toomre, J., Deane, A. E., Hurlburt,
N. E., Glatzmaier, G. A., Fichtl, G. H., Leslie, F., Fowlis, W. W.,
Gilman, P., 1986, Science, 234, 4772, 61

\bibitem[\protect\citeauthoryear{Miesch et al.}{2006}]{mbt06}Miesch, M. S., Brun, A. S., Toomre, J., 2006, ApJ, 641, 618

\bibitem[\protect\citeauthoryear{Miesch et al.} {2009}]{mt09}
Miesch, M. S., Toomre, J., 2009, Ann. Rev. Fluid Mech., 41, 317

\bibitem[\protect\citeauthoryear{Parfrey \& Menou}{2007}]{pm07} Parfrey, K. P., Menou, K., 2007, ApJ, 667, L207                  
\bibitem[\protect\citeauthoryear{Pedlosky}{1987}]{p87}
Pedlosky, J., 1987, Geophysical Fluid Dynamics, (Springer-Verlag: New York)

\bibitem[\protect\citeauthoryear{Roxburgh}{2001}]{r01} Roxburgh, I. W., 2001, A\&A, 377, 688

\bibitem[\protect\citeauthoryear{Schwarzschild}{1958}]{s58}
Schwarzschild, M. 1958, Structure and Evolution of the Stars (Dover: New York)

\bibitem[\protect\citeauthoryear{Stix}{2004}]{s04} Stix, M., 2004, The Sun, an Introduction (Springer-Verlag: Berlin)
\end{thebibliography}
\end{document}